\documentclass[12pt,a4paper]{article}
\usepackage{amsmath}
\usepackage{amssymb}
\usepackage{amsfonts}
\usepackage{array}
\usepackage{graphicx}% use this package if an eps figure is included.
\usepackage{mathrsfs}
\usepackage{multirow}
\usepackage{siunitx}
\newcommand{\bea}{\begin{eqnarray*}}
\newcommand{\eea}{\end{eqnarray*}}
\newcommand{\beao}{\begin{eqnarray}}
\newcommand{\eeao}{\end{eqnarray}}

\usepackage{authblk}
\makeatletter
\renewcommand\AB@affilsepx{, \protect\Affilfont} % Change separator to a comma
\makeatother

\usepackage{amsmath,amssymb,amsfonts}
\usepackage{amsthm}
\usepackage{xcolor}
\usepackage{nicefrac}
\usepackage{enumitem}
\usepackage{hyperref}
\usepackage[square,numbers]{natbib}
\setcitestyle{numbers}

\DeclareSymbolFont{symbolsC}{U}{txsyc}{m}{n}
\DeclareMathSymbol{\strictif}{\mathrel}{symbolsC}{74}
\usepackage{xcolor}

\title{Large Physics Models: Towards a collaborative approach with Large Language Models and Foundation Models}
\author[1]{Kristian G. Barman\thanks{Corresponding authors: Sascha Caron (\href{mailto:scaron@nikhef.nl}{scaron@nikhef.nl}) and Kristian G. Barman (\href{mailto:kristiancampbell.gonzalezbarman@ugent.be}{kristiancampbell.gonzalezbarman@ugent.be})}}
\author[2]{Sascha Caron$^*$}
\author[3]{Emily Sullivan}
\author[4]{Henk W. de Regt}
\author[5]{Roberto Ruiz de Austri}
\author[6]{Mieke Boon}
\author[7]{Michael F\"{a}rber}
\author[8]{Stefan Fr\"{o}se}
\author[9]{Faegheh Hasibi}
\author[10]{Andreas Ipp}
\author[11]{Rukshak Kapoor}
\author[12]{Gregor Kasieczka}
\author[13]{Daniel Kosti\'{c}}
\author[14]{Michael Kr\"{a}mer}
\author[15]{Tobias Golling} 
\author[16]{Luis G. Lopez}
\author[17]{Jesus Marco}
\author[18,19]{Sydney Otten}
\author[1]{Pawel Pawlowski}
\author[20]{Pietro Vischia}
\author[1]{Erik Weber}
\author[21]{Christoph Weniger}

\affil[1]{CLPS - Centre for Logic and Philosophy of Science, UGent, BE}
\affil[2]{IMAPP, Radboud University and Nikhef, NL}
\affil[3]{Utrecht University, NL}
\affil[4]{Institute for Science in Society, Radboud University, NL}
\affil[5]{Instituto de F\'{i}sica Corpuscular (IFIC), CSIC-UV, Spain}
\affil[6]{University of Twente, The Netherlands}
\affil[7]{TU Dresden \& ScaDS.AI, Germany}
\affil[8]{ErUM-Data-Hub \& TU Dortmund University, Germany}
\affil[9]{Computing and Information Science, Radboud University, NL}
\affil[10]{TU Wien, Austria}
\affil[11]{Thapar Institute of Engineering \& Technology (TIET), Patiala, India}
\affil[12]{Universit\"{a}t Hamburg, Germany}
\affil[13]{Institute of Philosophy, University of Leiden, NL}
\affil[14]{RWTH Aachen University, Germany}
\affil[15]{University of Geneva, Switzerland}
\affil[16]{Munich Center for Mathematical Philosophy, LMU Munich, Germany}
\affil[17]{Institute of Physics of Cantabria (IFCA), CSIC-UC, Spain}
\affil[18]{Radboud University, NL}
\affil[19]{Ippen Digital, Germany}
\affil[20]{Universidad de Oviedo and ICTEA, Spain}
\affil[21]{GRAPPA, Institute of Physics, University of Amsterdam, The Netherlands}

\renewcommand\Affilfont{\small} % Adjust affiliation font size

\begin{document}

\maketitle

\begin{abstract}
This paper explores ideas and provides a potential roadmap for the development and evaluation of physics-specific large-scale AI models, which we refer to as Large Physics Models (LPMs). These models, based on foundation models such as Large Language Models (LLMs) - trained on broad data - are tailored to address the unique demands of physics research. LPMs can function independently or as part of an integrated framework. This framework can incorporate specialized tools, including symbolic reasoning modules for mathematical manipulations, frameworks to analyse specific experimental and simulated data, and mechanisms for synthesizing insights from physical theories and scientific literature.
We begin by examining whether the physics community should actively develop and refine dedicated models, rather than relying solely on commercial LLMs. We then outline how LPMs can be realized through interdisciplinary collaboration among experts in physics, computer science, and philosophy of science. To integrate these models effectively, we identify three key pillars: Development, Evaluation, and Philosophical Reflection. Development focuses on constructing models capable of processing physics texts, mathematical formulations, and diverse physical data. Evaluation assesses accuracy and reliability through testing and benchmarking. Finally, Philosophical Reflection encompasses the analysis of broader implications of LLMs in physics, including their potential to generate new scientific understanding and what novel collaboration dynamics might arise in research. Inspired by the organizational structure of experimental collaborations in particle physics, we propose a similarly interdisciplinary and collaborative approach to building and refining Large Physics Models. This roadmap provides specific objectives, defines pathways to achieve them, and identifies challenges that must be addressed to realise physics-specific large scale AI models.
\end{abstract}

\section{Introduction}
Traditionally, Machine Learning models in physics were narrowly focused and designed for specific tasks such as distinguishing signal events from background events in particle accelerator experiments, predicting the mass of a particle given detector data, discovering materials, or identifying celestial objects
\cite{Albertsson:2018maf,Radovic2018,Boehnlein:2021eym,Cuoco:2020ogp,schwartz2021modern, calafiura2022artificial, hepmlc2024}. While some of these models were (and still are) effective in performing the specific tasks they were designed for, they lacked the versatility to be applied beyond their original domains. In contrast to domain and application specific models, Large Language Models (LLMs) such as GPT-4 ~\citep{achiam2023gpt}, Claude \cite{claude}, Gemma~\citep{team2024gemma} and Llama \citep{dubey2024llama}, are versatile and able to analyse and react to text, images, computer code, and data in general, across various domains, and with remarkable proficiency~\citep{li2024sora, yang2024video}.

Unlike narrow models, which are primarily used in data analysis, LLMs can enhance a broad range of research activities. Scientific research can be regarded as an interconnected network of processes aimed at advancing scientific understanding. In this network, activities such as hypothesis generation, experimentation, data analysis, and model or theory development continually feed back into one another in an iterative, dynamic fashion. The successful integration of LLMs into these workflows requires not only a careful evaluation of the skills they bring to such activities but also of their alignment with desirable epistemic values (e.g., accuracy, coherence, and explanatory power) and non-epistemic values (e.g., societal impact and ethical considerations). As was the case with the use of computer-based research, LPMs could also help us overcome the limits of our human capabilities. For instance, they may help speed up scientific progress in general, deepening our scientific understanding, which is one of the ultimate goals of science, enabling us not only to make accurate predictions but also to grasp why things are the way they are~\citep{deregt2017understanding}.

For instance, they can aid the generation of hypotheses by serving as brainstorming partners, and in doing so may offer novel approaches to complex problems in physics and inspire creativity, acting as an 'artificial muse'\citep{krenn2022scientific} for researchers. For example, \cite{gu2024interesting} use LLMs to suggest ideas in physics, subsequently having 100 experienced researchers evaluate the resulting ideas. In the experimental phase, LLMs may assist in devising tests and experiments, showcasing both current capabilities and future potential for experiment design and execution, thereby streamlining the research process~\citep{adesso2023towards}. In data analysis, LLMs can offer new methods for interpretation and visualisation, particularly in data-intensive physics. They could also generate insights that may elude human researchers or narrower AI models and automate code generation for data analysis~\citep{ali2023physics, otten2023learning, nascimento2024lm4ds}. Moreover, LLMs may help formulate accessible conclusions for a broader scientific community, facilitating knowledge sharing across disciplines, while also enabling multimodal integration and generalisation~\citep{zolna2024gats}.

To date, there have been notable examples of successful scientific use of LLMs, including the design and execution of autonomous experiments, agent-based AI labs, applications to symbolic tasks, and the development of collaborative AI tools known as CoScientists, and the enhancement of capabilities in programming and mathematical proof generation~\citep{swanson2024virtual, boiko2023autonomous, lample2019deep, huang2024crispr, madani2023large, romera2024mathematical, cai2024transforming}. It is claimed that LLMs are transforming scientific research by creating a 'hybrid intelligence' that augments human cognition and reshapes research processes~\citep{akata2020research, babushkina2022epistemo}. Despite LLM's limitations in reasoning and creativity~\citep{messeri2024artificial, joshi2024llms}, which are usually central to the process of hypothesis formulation, experimentation, and scientific conclusions, they have been shown to enrich the initial stage of scientific inquiry by posing insightful questions and providing valuable observations. In selecting relevant research topics, they can efficiently summarise and highlight key findings from vast scientific literature, and enhance understanding~\citep{bubeck2023sparks, bolanos2024artificial}. Additionally, LLMs facilitate the exploration of interdisciplinary research, bridging gaps between different fields and advancing innovative ideas.

Given the prospective success of LLMs, a fundamental question that arises is whether the physics community should use commercially available tools or develop a custom-built LLM specifically for physics research. While tools such as GPT-4 are frequently utilised in physics, their usefulness depends on accommodating a tool that was not primarily designed for physicists, and therefore requires either fine-tuning on additional data, or techniques such Retrieval-Augmented Generation (RAG)\citep{lala2023paperqa, soudani2024fine} (see ATLASchatbot - chATLAS). For instance, tools like AstroLLama\citep{nguyen2023astrollama} are fine-tuned on existing LLMs using abstracts of astronomy papers. Such tools, however, still have very limited capabilities, as fine-tuning on domain-specific data often lacks the comprehensive contextual understanding and reasoning skills necessary for addressing complex, novel problems in physics and astronomy. Similarly, there is a growing number of tools for more general physical data embedding that can solve very specific physical problems via a common physical embedding. These tools are often referred to as "foundation models" in the physics literature, but usually have no text interface (see e.g.,~\citep{parker2024astroclip, golling2024masked, Otten:2019hhl, harris2024simulation, birk2024omnijet,rizhko2024astrom}), and their scope is limited to very domain-specific data analysis problems and not widely generalisable. Additionally, recent results \citep{brehmer2024} suggest that combining physics-informed architectures with fine-tuning on specific tasks may provide the best performance - these architectures would encode fundamental physical principles while finetuning would allow adaptation to particular use cases.

After considering the arguments for and against, we propose that the physics community should develop and evaluate its own model of physical language. This initiative aligns with the community's fundamental mission of advancing physical understanding. While commercial AI models exist, the physics community possesses unique advantages that make them especially well suited to undertake this endeavor, namely, their domain expertise, access to specialized experimental data, and capability to conduct and validate physical experiments which produce novel data (see Section 2). We envision this model as an integrated framework comprising a network of dedicated foundation models \citep {bommasani2021opportunities} and Large Language Models (LLMs). We refer to these general-purpose, large-scale AI systems as Large Physics Models (LPMs). These AI models are tailored for physics research, designed to analyze, understand, and generate explanations of physical phenomena by processing formulas, relationships, and—via dedicated interfaces—experimental and simulated data. These models leverage the capabilities of LLMs to achieve broad multimodal understanding, allowing them to interpret and generate text, mathematical equations, diagrams, and data visualisations commonly used in physics. Such LPMs can have the potential to augment the fundamental structure of scientific inquiry~\citep{ai4science2023impact}. We foresee LPMs as a collection of interconnected LLMs and foundation models, each specialized for distinct domains within physics (see section 4). Interconnection can take place via central agents (routers) who select the specialized foundation models or LLMs according to their capabilities, providing a unified interface.

The overarching aim of this paper is to establish a potential roadmap for the development and evaluation of LPMs in the context of physics, while also reflecting philosophically on their transformative potential for scientific understanding. The development of an LPM represents an ambitious step forward in the application of AI in scientific practice.

The origin of this paper stems from a workshop held in early 2024 in Leiden, which brought together leading researchers in physics, computer science, and philosophy of science to discuss and explore the unique challenges and opportunities presented by LLMs with respect to scientific understanding in fundamental physics.

The paper is structured as follows: Section 2 discusses whether the physics community should develop LPMs, Section 3 outlines the three main pillars (Development, Evaluation, and Philosophical Reflection) and discusses the interactions among their respective communities. Section 4 details the objectives, challenges, and methods involved in developing an LPM. Section 5 examines the objectives, methods and challenges in evaluating and benchmarking an LPM. Section 6 considers the objectives of, and challenges and methods for philosophical reflection on the broader implications of AI in scientific research.

\section{Does the physics community need specialised large scale physics-specific AI models?}

The integration of LLMs and foundation models into physics research appears to be a promising and inevitable development. However, it raises the question of whether the physics community should actively develop and refine these AI models or rely on commercial ones. On the one hand, large commercial LLMs offer several advantages. These models are typically well-funded, benefiting from the vast computational resources and specialised AI expertise found in industry~\citep{ruan2024observational}. Commercial LLMs are also designed to be versatile and adaptable, potentially facilitating their application across various subfields of physics. Moreover, relying on commercial models could allow physicists to focus on their core research rather than diverting time and resources to AI development.

On the other hand, there are strong reasons for the physics community to take an active role in developing LLMs tailored to their specific needs. Physics research often involves complex mathematical formulations, symbolic reasoning, and domain-specific knowledge that may not be adequately captured by general-purpose commercial models~\citep{zhang2024chemllm, hudson2023trillion}. By developing models in house, the physics community can better align these tools with the unique methodologies, standards, and methodological considerations of their field~\citep{park2024ai, russo2023connecting, sharma2023towards}. This level of control and customization could lead to more accurate, interpretable, and trustworthy results~\citep{zou2023representation}. Moreover, maintaining control over the training data, model architectures, and evaluation processes, the community can ensure adherence to scientific principles such as reproducibility and peer review~\citep{duede2023deep, messeri2024artificial, zenil2023future}.

Weighing these considerations, we believe that the ideal scenario would be for the physics community to take a leading role in developing LLMs and foundation models tailored to their needs. This approach would allow for the creation of AI tools that are finetuned to the complexities of physics research, aligned with the community's values and standards, and optimised for advancing fundamental knowledge and understanding.

We acknowledge that this ideal scenario faces significant challenges, particularly in terms of funding and resources. Developing and maintaining cutting-edge AI models requires substantial computational power, specialised expertise, and ongoing investment. To mitigate these challenges, we propose a multifaceted approach. The physics community should actively seek out strategic collaborations with industry partners. By leveraging the computational resources and expertise of commercial companies while maintaining a leadership role in model development, the community can see to it that the resulting tools are tailored to its needs while benefiting from the scale and capabilities of industry. Furthermore, the community should explore innovative funding models, such as targeted grants, partnerships with philanthropic organisations, support by large international research organizations or laboratories (e.g. CERN) and cross-disciplinary collaborations that pool resources and expertise.

Additionally, adopting a collaboration structure similar to those used in particle physics for experimental analysis may provide an effective framework for building physics-specific AIs. Such a structure promotes the pooling of resources, the sharing of expertise and involvement across institutions, and the coordination of efforts on a large scale, which are all crucial for tackling the complex challenges associated with developing large-scale AI models. The development of open-source platforms and shared infrastructure could help spread the costs and foster a collaborative ecosystem around physics-specific LLMs.

While the rapid pace of commercial AI development creates a "dead on arrival" risk - where physics-specific models could become obsolete before completion due to faster industry development cycles - several factors make this challenge manageable. The physics community possesses unique advantages that create a natural moat protecting the enduring value of specialized models: domain expertise in fundamental physics, control over specialized datasets and on the ability to create new data (e.g. performing experiments), and development of domain-specific software and simulations that cannot be readily replicated by external organizations without collaboration. These strategic assets are particularly valuable because many physics problems cannot be simply solved by increases in computational power.  Moreover, there are compelling reasons to maintain a certain degree of independence from commercial AI, including data privacy concerns and the need for transparent, ethically-aligned models that adhere to scientific standards. To maintain relevance and competitiveness, we advocate for an adaptive development strategy that combines: (1) iterative, modular development allowing continuous updates, (2) carefully scoped open-source collaboration enabling community-wide contributions in certain areas, (3) transfer learning from existing models to accelerate development, and (4) strategic partnerships that facilitate rapid integration of new advances while preserving the community's autonomy and scientific integrity.

An AI model for physics can also serve as a prototype for other fields of science. The physics community has a long-standing tradition of collaboration on large-scale projects, shared data repositories, and rigorous peer review processes and is experienced in addressing challenges similar to those encountered in large AI projects (see e.g.~\cite{engelen2012large}). Furthermore, physicists' work in theoretical and experimental contexts offers diverse testing grounds for evaluating AI capabilities in simulation, pattern recognition, and hypothesis testing. This foundational experience could streamline the adaptation of similar AI architectures across other scientific domains. The next question that arises then is exactly how can we obtain an LPM.

\section{How to get an LPM: Key Pillars}
To address the challenges in developing LPMs and to harness their potential for advancing physics, we propose a roadmap structured around three critical pillars: Development, Evaluation, and Philosophical Reflection. This roadmap is designed specifically for the physics community, but should in principle be applicable across other scientific disciplines.

The Development pillar focuses on creating robust LPMs capable of handling the complexities of physical theories, data, and natural language. This involves interdisciplinary collaboration to safeguard that the models are not only powerful but also well-suited to the specific needs of physical research. Key objectives within this pillar include: (i)~developing foundational models tailored to physics, (ii)~curating high-quality, diverse physics datasets, (iii)~integrating physics-specific knowledge and reasoning capabilities, (iv)~enabling interaction with physical databases and simulators, and (v)~continuously updating models to keep pace with scientific progress. Underlying all these is perhaps the most important component: (vi)~the development of collaboration platforms.

The Evaluation pillar is dedicated to assessing the accuracy, reliability, efficiency, and effectiveness of LPMs, which involves testing and benchmarking. Evaluators play an important role in validating the models, improving the trustworthiness and scientific soundness of their outputs. Key focus areas within this pillar include: (i)~developing physics-specific benchmarks and evaluation protocols, (ii)~assessing model accuracy on complex physics tasks, (iii)~verifying reliability and robustness across diverse scenarios, (iv)~evaluating the practical utility of models for real-world physics research, and (v)~enabling integration into existing scientific workflows.

The Philosophical Reflection pillar involves examining the broader implications of integrating LPMs into scientific practice. This includes exploring their potential to generate new scientific insights, transform traditional problem-solving approaches in physics, and reshape the nature of scientific understanding itself. Philosophers of science address fundamental questions such as: (i)~Can LPMs truly possess scientific understanding, or are they merely powerful tools for human researchers? (ii)~How do we define and measure scientific understanding in the context of AI? (iii)~What are the epistemological and ethical implications of relying on AI models in scientific discovery? (iv)~How can we ensure that the use of LPMs aligns with the values and goals of scientific inquiry?

The roadmap emphasises the interconnected nature of these pillars, highlighting the importance of ongoing dialogue and feedback loops between developers, evaluators, and philosophers of science. As depicted in Figure~1 below, these three communities—developers, evaluators, and philosophers of science—must work in close collaboration to realise the full potential of LPMs. Developers create models that are informed by the insights of evaluators and philosophers, aiming towards making the models not only powerful but also reliable, interpretable, and aligned with the needs of physics research. Evaluators, in turn, provide rigorous assessments of model performance and help identify areas for improvement, while also considering the broader implications highlighted by philosophers. Philosophers of science contribute conceptual clarity, epistemological guidance, and frameworks that shape the development and deployment of LPMs, seeing to it that these tools remain grounded in the fundamental principles of scientific inquiry.

\begin{figure}[htbp]
    \centering
    \includegraphics[width=0.8\textwidth]{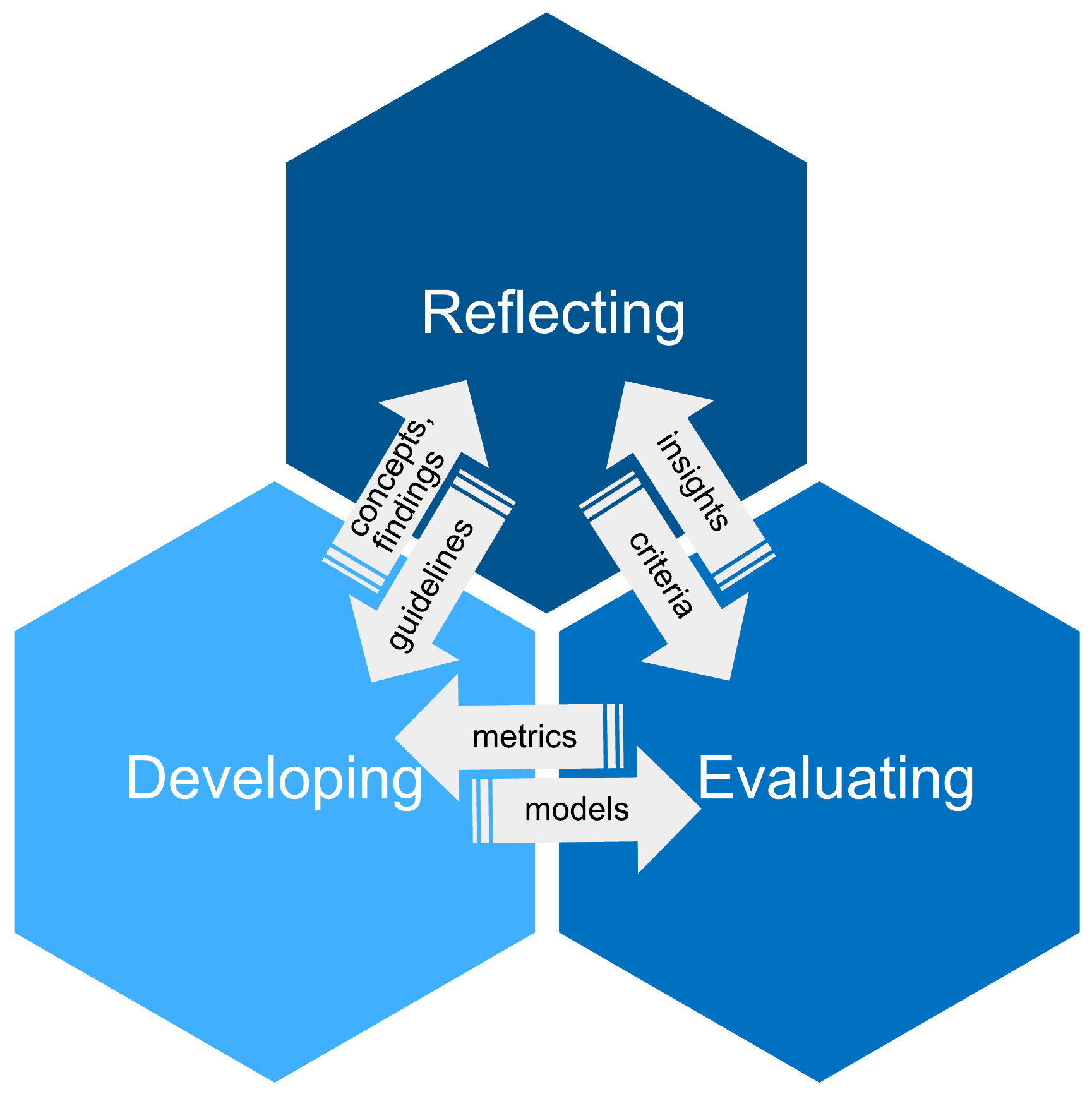}
    \caption{\textbf{Dynamic Interactions Between the Three Pillars}. This diagram depicts an interconnected and cyclical system comprising three pillars - Developing, Evaluating, and Reflecting - that operate through continuous exchanges. The system functions through input-output relationships between each pillar: between Developing and Evaluating, where Developing generates models and receives metrics; between Evaluating and Reflecting, where Evaluating produces insights and receives criteria; and between Reflecting and Developing, where Reflecting provides guidelines while incorporating concepts/findings. These exchanges create a dynamic framework where each pillar both influences and is influenced by the others, enabling systematic improvement through continuous feedback loops and iterative evolution. 
}
\end{figure}

\section{Development Pillar}

The development pillar focuses on the construction of LPMs that can navigate the complex landscape of physical theories, physical simulators, and real-world experimental conditions while simultaneously mastering science-specific tasks such as symbolic reasoning, hypothesis generation, and the interpretation of complex data sets.

\subsection{Objectives}

The primary goal is to develop foundation and LLM models tailored to the needs of science (see~\cite{hudson2023trillion}, in chemistry see~\cite{zhang2024chemllm,ai4science2023impact,bran2024augmenting,boiko2023autonomous}). In the context of physics, this involves pretraining on physics data and fine-tuning using curated datasets that comprise scientific papers, textbooks, and problems from physics. Consequently, this process should aim at teaching the model the language nuances, theories, and problem-solving strategies specific to physics.

To be of value to the physics community, models must be equipped with  the ability to analyse experimental data, perform simulations, and compare simulated data with experimental physical data~\citep{liu2024world}. Equally, this requires writing domain-specific computing code and accessing mathematical and statistical software tools or the ability to access other foundation models tailored to the analysis of specific experimental data. Since mathematics is the language in which physics underpins its theories and principles, close collaboration with mathematicians and symbolic computation experts will be valuable. This cooperation can guide the model's ability to process and generate symbolic representations, perform algebraic manipulations, and apply advanced mathematical methods such as calculus, linear algebra, and tensor operations. Here, collaboration with the symbolic math in AI community will be an important step to advance physics AI models (see, e.g., \cite{mathai2024,ai4math2024}.
 
Given the wide range of subfields within physics, it will be necessary to develop models that meet the specific requirements of each domain. Incidentally, this could involve creating  specialized multi-purpose foundation models for particle physics, astrophysics, condensed matter physics, and other areas, each trained on domain-specific data and equipped with the necessary knowledge and problem-solving capabilities, with a common pre-training and subsequent specialised modules. For instance, foundation models tailored to particle physics could focus on tasks such as event classification, detector simulation, or reconstructing collision events, leveraging the unique data generated by particle accelerators. Similarly, an astrophysics-focused model~\citep{alvarez2022snowmass2021} could analyze large-scale cosmological simulations, process observational data from telescopes, or assist in interpreting phenomena like gravitational waves and exoplanetary systems. As a first step, building small-scale demonstrators for specific tasks within each subfield would provide proof of concept, allowing researchers to evaluate feasibility, performance, and potential impact. Recently, preliminary models for particle physics~\citep{wildridge2024bumblebee} and heavy-ion collision experiments~\citep{kuttan2024towards} have been proposed, demonstrating early approaches and ideas towards foundation models that can be pretrained and fine-tuned or used in versatile ways. The domain-specific models would not only enhance performance by concentrating on the unique challenges of each field but could also be integrated into a larger, interdisciplinary framework, enabling cross-domain insights and collaborative problem-solving (see Figure~2).

\begin{figure}[htbp]
    \centering
    \includegraphics[width=0.95\textwidth]{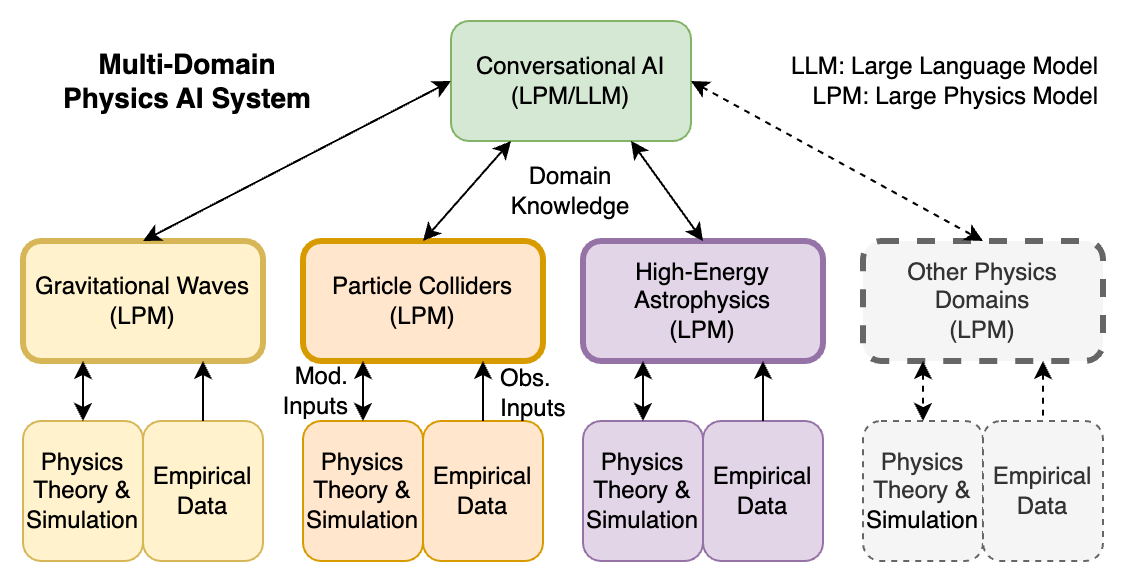}
%    \caption{\textbf{Modular Architecture of LPMs}.  The figure depicts the hierarchical organisation of a LPM, where a central Text model integrates specialised models tailored for distinct subfields, including Gravitational Wave Physics (GW), Large Hadron Collider (LHC) data, and Astrophysics.}
    \caption{\textbf{Large Physics Models (LPMs) as central components of a Multi-Domain Physics AI System}.  The figure depicts the central role that tailored LPMs play in connecting conversational AI models (LLM trained as LPM) with the theoretical, computational and empirical knowledge and infrastructure of distinct physics subfields, including
%    }
    %where a central Text model integrates specialised models tailored for distinct subfields, including 
    Gravitational Wave Physics, Particle Colliders such as the Large Hadron Collider (LHC), and high-energy astrophysics. Subfield LPMs are guided by a conversational AI with embedded physics knowledge.
    }
\end{figure}
In particular, we envision future scientific research being enhanced by the concept of an 'AI physicist,' a system of interconnected foundation models tailored to tackle specific research tasks across various domains of physics~\citep{otten2023learning}; \cite{durante2024interactive},p. 20). This strategy draws inspiration from the division of labour in collaborative research teams, aiming to increase productivity and innovation in physics research. This is also inline with recent advancements in the orchestration of multiple LLM to perform complex question answering tasks ~\citep{hoveyda2024aqa,zhugegptswarm}. Accordingly, the network encompasses specialised foundation models (e.g., trained on the corpus of open-access peer-reviewed journals, on the data of the Large Hadron Collider (LHC), Gravitational Wave Physics (GW), and Astrophysics) that handle literature synthesis, data analysis and simulations, result visualisation, and scientific paper composition, covering the full spectrum of research activities. In this context, agents are autonomous AI systems that can perceive their environment, make decisions, and take actions to achieve specific goals. These agents can operate in coordination, supported by APIs and frameworks for seamless information exchange, while human researchers use the models and guide and refine the process, improving alignment with genuine scientific inquiry objectives.

Ideally, (conversational) LPMs should be capable of generating new hypotheses, proposing innovative experiments or questions to pose to the corpus of data, and identifying promising research directions. Through processing vast amounts of scientific literature and data, these models may uncover hidden patterns, connections, and gaps in existing knowledge, thereby guiding researchers towards novel discoveries and breakthroughs. Notwithstanding, to facilitate effective collaboration between human researchers and LPMs, it is important to develop user-friendly interfaces (presumably through natural language, but perhaps also incorporating key visual components) that allow scientists to interact with the AI seamlessly. These interfaces should enable researchers to input their queries, provide guidance, and interpret the model's outputs easily. In this way, by designing intuitive interfaces, the development pillar aims to bridge the gap between the technical aspects of AI and the domain expertise of physicists, enabling a synergistic partnership.
\subsection{Challenges and Methods}

The development of physics-specific large-scale AI models presents a unique set of challenges, ranging from data curation and processing to model design, high-performance computing (HPC) and evaluation. These challenges are closely interrelated and some require innovative approaches and interdisciplinary collaboration to meet the diverse needs of physics research. In what follows, we briefly discuss some of these challenges and outline initial approaches to address them. While these efforts provide a basic perspective, further extensive work is required to fully address these issues, particularly through the development of initial demonstrator models.

Curating high-quality, diverse datasets is a significant challenge in developing LPMs, especially when handling experimental and simulated data. These data sources often come in various formats and require careful preprocessing to ensure compatibility with AI models. Collaboration with benchmark developers (evaluators) will be necessary not only in standardizing data formats, but also in creating evaluation frameworks. These benchmarks (described in the next section) help evaluate model performance on various physics-specific tasks, provide a basis for comparison between different approaches and guide the iterative improvements during development.

When developing physics-specific large-scale AI models, an important question will be whether existing open-source models should be adopted or completely new models tailored to physics tasks should be created. A pragmatic strategy could involve a hybrid approach that leverages the strengths of both methods. First, the physics community could fine-tune open source base models for specific use cases, such as processing scientific literature or solving domain-specific problems. This would enable rapid prototyping and early demonstrations to showcase the feasibility of LPMs while minimizing resource requirements. Over time, as expertise and resources increase, the focus could shift to developing specialized models optimized for physics-specific challenges, such as symbolic reasoning, mathematical problem solving, and experimental data analysis. Methods such as Retrieval-Augmented Generation (RAG) could enhance LPMs by incorporating external knowledge into the prompt, expanding the model's reach beyond its internalised data~\citep{lala2023paperqa,soudani2024fine}). This method retrieves relevant information from structured knowledge bases and scientific literature, producing more accurate and detailed responses to scientific queries. Usually a vector database is created from embeddings of documents but more advanced approaches are also feasible (e.g., constructing a graph from the documents- GraphRAG).

In parallel to developing LPMs, domain-specific data processing pipelines (domain-specific foundation models) should be developed to handle the unique formats and complexities of physics data, such as collider datasets from the Large Hadron Collider (LHC) or astrophysical observations. These pipelines play a critical role in preprocessing, standardizing, and integrating diverse data types into AI workflows, which will be needed in order to have compatibility and coherence across different modalities.
Such pipelines will initially rely on the existing analytics infrastructure (e.g. within experimental collaborations or theoretical simulations) but will progressively incorporate a combination of end-to-end machine learning and trainable physics-based algorithms.

One of the key challenges will be determining how to represent physics data so that general-purpose LLMs can operate effectively. Tokenization strategies\cite{leigh2024tokenization} may be particularly useful for encoding physics-specific data, such as particle sequences, event records, mathematical equations, and simulation outputs, into formats suitable for LLMs. For instance, pretraining models on raw experimental data, like LHC particle sequences or astrophysical time-series datasets, may enable them to learn domain-specific patterns and structures, improving their ability to interpret new experimental data and to uncover underlying physical phenomena. However, to fully exploit the richness of physics data, alternative approaches besides tokenization - such as continuous representations, graphs or neural fields - should be explored, which may provide more seamless and meaningful representations of physical structures.

In addition to tokenization, the creation of generic physics data embeddings will also play a key role. These embeddings could encode experimental and simulated information in a standardized, LLM-compatible format, allowing for broader applicability across subfields. This effort requires integrating diverse data types, including text, images, mathematical formulas, and experimental results, into a coherent training pipeline. Achieving this integration will pose a significant challenge but will be a prerequisite for building versatile LPMs.

Furthermore, connecting LPMs to external resources, such as databases like the Particle Data Group (PDG)~\citep{navas2024review} or physics-specific simulation software, could enable real-time analysis (and re-analysis) and facilitate comparisons between experimental and simulated data. Such integration would not only enhance model accuracy but also provide tools to refine theoretical models, bridging the gap between theory, experiment, and AI-driven insights.
Early attempts to automate the search for new physics and the re-analysis of data in particle physics at the Tevatron, LEP, HERA and LHC colliders, for example, are \cite{D0:2001jkr,Cranmer:2010hk,Holzner:2009ew,Caron:2006fg,D0:2000vuh,Abbott:2000gx,Abbott:2001ke,Abazov:2011ma,H1:2004rlm,H1:2008aak,CDF:2007iou,CDF:2008voc,Aaboud:2018ufy,Sirunyan:2020jwk}.

Another significant challenge in developing LPMs lies in symbolic reasoning and the manipulation of complex mathematical formulas. Techniques such as Abstract Syntax Trees (ASTs), theorem provers, and symbolic algebra systems have shown great promise for enabling models to parse and manipulate mathematical structures~\citep{pantsar2024theorem}. ASTs represent mathematical expressions hierarchically, aiding in equation parsing, while theorem provers allow for logical reasoning and inference which will likely be useful for physics problem-solving.

Additionally, the neuro-symbolic AI approach, which merges neural networks' pattern recognition capabilities with symbolic reasoning engines, also holds promise for LPMs~\citep{karpas2022mrkl,cunnington2024role}. This hybrid method allows models to both process language and reason about equations, making it particularly effective for complex scientific tasks.

Training data inconsistencies, biases, and errors pose additional challenges
~\citep{pagano2023bias,shah2024comprehensive}. For instance, historical biases in scientific literature or limitations in experimental datasets can introduce skewed model outputs. Here, Explainable AI (XAI) methods~\citep{zhao2023explainability}, such as attention visualisation and feature attribution, may help improve some of the key aspects of the transparency of LPMs by revealing the key features and reasoning steps driving certain relevant models' outputs. XAI methods may be augmented through 'mechanistic interpretability,' which encompasses interventionist approaches~\citep{nanda2023progress,conmy2023towards,geiger2024causal}  that are expected to shed light on the internal causal mechanisms driving model behaviour~\citep{milliere2024interventionist} and offer insights into the broader functional structures that connect these mechanisms to output generation~\citep{kastner2024explaining}, thus enhancing transparency and reliability. While these methodologies are still in a nascent phase, they are increasing in sophistication at a rapid rate.

As LPMs become increasingly integral to research, ensuring their long-term accessibility and support—beyond commercial interests—becomes another consideration. Open-source development, community-driven maintenance, and dedicated funding initiatives are vital strategies to safeguard the models' availability for future scientific advancement.This brings up the issue as to why researchers will decide to adopt LPMs. The reasoning lies in a game-theoretic incentive: researchers who adopt LPMs will gain a competitive advantage in their ability to generate novel insights and accelerate their research progress, creating pressure for adoption.

Lastly, developing large-scale LPMs requires substantial high-performance computing (HPC) infrastructure, which can be prohibitive for some research groups. Collaborative efforts with HPC centres or cloud computing providers, along with the development of more efficient computational solutions, will be needed to democratise access to these powerful models within the scientific community.
\section{Evaluation Pillar}
The Evaluation Pillar is tasked with assessing the accuracy, reliability, and effectiveness of LPMs. Through testing and benchmarking, evaluators play a fundamental  role in validating the models and contributing to making their outputs more trustworthy and scientifically sound.

\subsection{Objectives}

The primary objective of the Evaluation Pillar is to assess LPMs' capabilities, such as scientific reasoning and discovery. Accordingly, evaluators must develop benchmarks that test core physics knowledge, mathematical abilities, and research aptitude. These benchmarks should cover a wide range of physics subfields and difficulty levels, from basic concepts to advanced research-level problems.

To develop robust and generalizable (conversational) LPMs, it is important to evaluate their performance under distribution shift and on out-of-domain problems~\citep{yuan2023revisiting,zhang2024chemllm,wang2024beyond}. The models are then tested on data sets and tasks that differ from their training data to assess their ability to adapt and apply their knowledge to new situations. Additionally, evaluators should measure model robustness to input perturbations, adversarial attacks, and corrupted data, ensuring that the models can handle noisy and imperfect inputs that may be encountered in real-world research settings.

Assessing model calibration and uncertainty quantification on experimental data analysis tasks is another critical objective of the Evaluation Pillar. LPMs should not only provide accurate predictions but also express appropriate levels of confidence in their outputs. Consequently, evaluators must develop methods to measure the alignment between model predictions and real-world physics data and theories, seeing to it that the models generate scientifically valid and reliable results.

Comparing the performance of LPMs to that of human experts on complex, research level challenges can help with understanding the extent to which these models can augment human capabilities. Equally, evaluators should design benchmarks that require a combination of deep physics knowledge, creativity, and problem-solving skills, pushing the boundaries of what AI can achieve in scientific discovery. By quantifying the efficiency gains and acceleration to scientific workflows enabled by LPMs, evaluators can demonstrate the practical value of these models in streamlining research processes and accelerating progress.

Investigating the theoretical underpinnings of artificial reasoning and understanding in LPMs is another important objective of the Evaluation Pillar. Evaluators should collaborate with philosophers of science and AI theorists to explore questions such as: What constitutes genuine scientific understanding in AI systems? How can we formalise and measure the reasoning capabilities of LPMs? Whereupon, by addressing these fundamental questions, evaluators can contribute to the development of a solid theoretical foundation for the use of AI in scientific discovery.

Tracking the performance improvements of LPMs over time in order to measure the rate of scientific progress enabled by these models presents another relevant task. Evaluators should establish longitudinal benchmarks that assess the models' capabilities across different versions, prompting scenarios, and training iterations, allowing researchers to identify trends, bottlenecks, and areas for further improvement. Incidentally, this information can guide the development of more advanced and efficient LPMs, accelerating the pace of scientific discovery.
\subsection{Challenges and Methods for Evaluating LPMs}

To date, there are many different benchmarks of physics for LLMs. Most of them focus on basic high-school level questions or focus on broader knowledge retrieval tasks~\citep{clark2022decentring,bisk2020piqa,arora2023have,wang2023scibench,schoenick2017moving}. However, there are no fundamental physics benchmarks, and there are no benchmarks that specifically target scientific understanding. Several methods are currently employed to assess LLMs in their capability to handle complex reasoning tasks, particularly in physics~\citep{rein2023gpqa,sawada2023arb,yue2024mmmu}. Theoretical approaches also contribute to this field; \cite{krenn2022scientific} suggests a scenario where the understanding of a teacher (AI or human) is evaluated based on their ability to transfer knowledge to a student, as judged by an independent referee. Similarly, the Scientific Understanding Benchmark (SUB) introduced by~\cite{barman2024towards} focuses on measuring scientific understanding through tasks involving information retrieval, explanation production, and the generation of counterfactual inferences. One could conceive of further hypothetical benchmarks. These might include the ability of LLMs to provide explanations that withstand scrutiny by domain experts or to create causally realistic simulations.

Current benchmarks for evaluating LLMs  (and future LPMs) present several challenges that evaluators need to address. A major issue is avoiding benchmark overfitting and Goodhart's law, which warns that when a measure becomes a target, it ceases to be a good measure~\citep{recht2019imagenet}. Therefore, evaluators must be cautious not to create benchmarks that are too narrow or easily gameable, designing diverse, comprehensive benchmark suites that test a wide range of skills, from conceptual understanding to creative problem-solving. General AI benchmarks often suffer from arbitrary task selection, incomplete domain coverage, and poor performance on minority sets (~\cite{raji2021ai}; see however~\cite{ruan2024observational}). Similarly, prompt-based evaluations can sometimes be brittle and sensitive to minor variations in the prompt wording. Physics-specific benchmarks must avoid these pitfalls. To achieve this, crowdsourcing challenge problems and evaluation from diverse members of the physics community could be useful as a way of developing meta-benchmarks that test generalisation and assess the trade-offs between breadth and depth of knowledge, as well as adversarial testing. Note however that there is a problem that benchmarks become part of the training data, therefore it is important that some benchmark data should not be published.   

Additionally, incorporating real experimental data, such as data from the Large Hadron Collider (LHC) or astrophysical images or simulated data, into LPM benchmarks might be useful for assessing LPMs' ability to reason about and analyse complex, multimodal scientific information. These real-world datasets often contain noise, artefacts, and other challenges that are not present in idealised benchmark problems.
Benchmarking large pre-trained models (LPMs) on domain-specific physical data, such as collider physics, astrophysics, or gravitational waves, using both real and simulated (labeled) datasets, is crucial for their effective deployment. Several benchmarks already exist in the literature, addressing tasks such as signal detection, reconstruction, and anomaly searches (e.g., \cite{Regimbau:2012ir,Adam-Bourdarios:2015pye,Aarrestad_2022,Kasieczka:2021xcg,Amrouche:2019wmx,MockLISADataChallengeTaskForce:2009wir,Eller:2023myr}). A detailed discussion of open data initiatives and specific benchmark developments, while important, is beyond the scope of this work.

Another key challenge is distinguishing memorization from reasoning. Evaluators should use methods like step-by-step explanation, counterfactual reasoning, out of distribution data, and process supervision to probe models' understanding. This can help models generalise beyond memorised information. Similarly, understanding (as well as other skills) is not a binary, but rather a gradient that may have many different levels. This means that a good benchmark should account for different types and different levels of understanding~\citep{barman2024towards}. Given the combinatorial space of physics problems, benchmarks must cover a representative sample to ensure models can handle diverse and novel problems. As scientific knowledge evolves, models must be updated to account for changing priors and assumptions. Thus, dynamic benchmarking frameworks are required to maintain relevance. Additionally, leaderboards and comparative benchmarks can drive short-term metric hacking and a narrow focus on beating the state-of-the-art. Evaluators should be aware of these potential negative impacts and prioritise long-term scientific progress and conceptual understanding over merely achieving the highest scores. 
%For instance, they might compare the performance of LPMs to theoretical upper bounds, such as Bayesian physics oracles, which can serve as a useful benchmark for assessing their efficiency and optimality~\citep{bradley2019bayes}.

Instituting data ablations and probing techniques can help dissect the sources of knowledge in LPMs and understand which lower-level components are responsible for certain behaviours~\citep{rosa2020measuring,na2024scalable}. Data ablations~\citep{meyes2019ablation} involve systematically removing or modifying specific subsets of the training data and observing the impact on model performance. Probing techniques involve designing targeted queries or tasks that test the models' understanding of specific concepts or principles. In using these, and other mechanistic interpretability techniques 
%(e.g. sparse autoencoders, causal scrubbing)%
, evaluators might gain a more fine-grained understanding of the models' knowledge and reasoning capabilities. Similarly, incorporating process supervision (as they work through a problem), such as their attention mechanisms or feature activations, into the evaluation framework could provide valuable insights into the reliability of the models' intermediate reasoning steps.

Lastly, implementing detailed error analysis frameworks could provide a systematic pathway to diagnose failure points in LPMs and identify areas for targeted improvement. For instance, through disaggregated error metrics~\citep{lu2023error}, researchers could decompose model performance across different problem types and difficulty levels, potentially revealing patterns of inconsistencies in physical reasoning. This granular understanding of failure modes could potentially illuminate specific weaknesses in LPM architectures, in addition to providing insights for developing more robust and theoretically-grounded models.

\section{Philosophical Reflection Pillar}

The integration of LPMs into scientific research presents a range of epistemological, conceptual, and ethical challenges that require philosophical investigation. The Philosophical Reflection Pillar focuses on the philosophical implications of integrating LPMs into scientific research. It explores how these AI systems might enhance or transform scientific practice, and reflects on their potential to increase scientific understanding, on the role of human scientists in their development and deployment, and on the types of collaborative frameworks that might be required.

\subsection{Objectives}

Examining the nature and criteria of scientific understanding in the context of AI presents a first objective. Whether, and if so how, LPMs can enhance human understanding is a philosophical question that relates to current debates on scientific understanding~\citep{khalifa2017understanding,deregt2017understanding,lawler2022scientific}. A subsequent issue is whether they can generate 'artificial understanding' independently of humans. While LPMs may support hypothesis generation and experimental testing, as, yet it is human scientists who integrate these results into broader understanding processes. In classical scientific practice, scientific understanding is generated by answering explanation-seeking questions of various types~\citep{weber2013scientific}, which are usually driven by epistemic interests such as foreseeing the consequences of interventions. The prospect of LPMs functioning as 'artificial scientists' who might possess 'artificial understanding' raises the question of whether such understanding should conform to extant criteria for human understanding or requires a novel conception of understanding~\citep{barman2024towards}. As LPMs evolve from tools to autonomous agents, in line with notions such as Zytkow's 'robot discoverer' \citep{zytkow1995creating}, the 'Robot Scientist'\citep{sparkes2010towards}, and the recent 'Intelligent Agent system'\citep{boiko2023autonomous}, their growing role in theory development and conceptual innovation is an area that needs to be investigated.

The integration of LPMs also introduces risks, such as the proliferation of unreliable research, reduced peer interaction, and the creation of echo chambers. These risks are compounded by the opacity in methodologies in complex fields, requiring the development of ethical guidelines and oversight mechanisms~\citep{park2024ai,russo2023connecting,sharma2023towards,barman2024beyond}. To address these challenges, philosophers must collaborate with AI researchers and physicists to develop frameworks for responsible and trustworthy AI in physics research.
\subsection{Challenges and Methods}

Defining and operationalizing \emph{scientific understanding} for AI systems, such as developing a clear and applicable definition, is a key challenge, as is the navigation of the tensions and trade-offs between explainability, accuracy, and complexity~\citep{beisbart2022philosophy,chirimuuta2021prediction,sullivan2022understanding}. The increasing autonomy of LPMs in scientific tasks raises concerns about their potential to replace human scientists and the implications for the role of creativity and serendipity in scientific discovery~\citep{tigre2023artificial}. While in some regards AI might present an opportunity for the long awaited '\emph{logic of discovery}'\citep{nickles1980scientific}; \citep{nickles1980introductory} that was sought after in the 70s and 80s, it appears the role of having humans in the loop is still very much needed\citep{zhou2022towards,he2024collaborative,kyriakou2023humans}.

Improving the alignment of AI with scientific values and ethical principles, such as objectivity, transparency, reproducibility, beneficence, non-maleficence, and justice, is another significant challenge. Similarly, the successful integration of LPMs into scientific research will require overcoming disciplinary silos and facilitating effective cross-disciplinary collaboration between physics, computer science, and philosophy. Achieving this transformation requires learning from successful experimental collaborations in physics and other fields, where clear incentive structures, shared infrastructure, and "win-win" scenarios enabled researchers to advance individual goals while contributing to broader objectives. This involves, among other things, aligning career incentives, funding mechanisms, and recognition systems with collaborative goals (see section below).

To address these challenges, philosophers would employ a range of methods. \emph{Conceptual analysis} and philosophical investigation, to clarify key concepts such as scientific understanding, artificial intelligence, and the nature of scientific discovery, and to explore the implications of LPMs for these concepts. \emph{Case studies} and historical examinations of the use of AI in scientific practice to provide insights into the evolving role of AI in scientific research. \emph{Thought experiments} and \emph{counterfactual reasoning} will be employed to explore the potential implications of LPMs for scientific understanding and discovery, and to test the boundaries of our concepts and frameworks. \emph{Ethical frameworks} and \emph{value alignment methodologies} will be developed to help steer in a direction such that the development and deployment of LPMs in scientific research is aligned with core scientific values and ethical principles. Finally, \emph{scenario planning} and \emph{anticipatory governance} methods could be used to explore the potential future trajectories of AI development and deployment in physics research, and to develop proactive strategies for managing the associated risks and challenges.

\section{Conclusion and Recommendations}

The development of large scale physics-specific AI models (LLM and foundation models), which we name Large Physics Models, would represent a significant step forward in the application of AI to scientific research. Creating AI systems that co-create physical theories, formulas, and relationships, might lead to generating novel hypotheses, experiments, and research directions. Therefore, LPMs could enhance both the pace of scientific discovery and the depth of our theoretical understanding. The proposed three-pillar framework—Development, Evaluation, and Philosophical Reflection—provides a structured approach to realizing this vision. 

We recommend the physics community to become active in the development of LPMs. This initiative should begin with small-scale demonstrator projects across specific physics subfields to validate approaches before scaling to larger efforts. To go beyond small scale demonstrators requires the  creation of larger-scale working groups dedicated to the development of LPMs. \textit{These efforts could evolve into formalized collaborations along the lines of experimental particle physics, with defined governance structures, shared computing resources and long-term support.}
 
We propose starting with a core group, who should establish governance and collaboration frameworks while developing initial demonstrations. Strategic industry partnerships will provide computational resources, while maintaining physics community leadership in model development. Funding should come through targeted grants and research institutions like CERN, with an aim at building cross-disciplinary collaborations that pool resources and expertise. To navigate the resource-intensive "valley of death" period, we recommend: setting clear milestones that demonstrate incremental value (while maintaining traditional research outputs), securing early partnerships with industry and funding bodies, and building community engagement through workshops, shared tools, and collaborative platforms. Success requires balancing short-term results  that keep the community engaged with the long-term vision of developing tailored LPM capabilities.

The successful development of LPMs could serve as a model for other scientific fields seeking to leverage AI capabilities, potentially accelerating scientific discovery and deepening our understanding while maintaining the fundamental principles that underpin good science.

\section{Acknowledgements}

We thank the Lorentz Center (Leiden University) for hosting and financially supporting the workshop ``Physics and Question-Answering Machines: Artificial Scientific Understanding?'', held in Leiden in January 2024, where this paper was conceived.

The work of Caron, De Regt, Barman was supported by an IRP grant from FNWI, Radboud University. Barman's work was also supported by FWO grant 1229124N.

Pietro Vischia's work was supported by the ``Ram\'on y Cajal'' program under Project No.\ RYC2021-033305-I funded by MCIN/AEI/10.13039/501100011033 and by the European Union NextGenerationEU/PRTR.

The work of R.\ RdA was supported by PID2020-113644GB-I00 from the Spanish Ministerio de Ciencia e Innovaci\'on and by the PROMETEO/2022/69 from the Spanish GVA.

Luis Lopez's work was partially supported by the Deutsche Forschungsgemeinschaft (DFG) under Grant No. 254954344/GRK2073/2.

The work of Pawel Pawlowski was supported by FWO grant 1255724N.

The views and opinions expressed are solely those of the authors and do not necessarily reflect those of the European Union or the European Commission. Neither the European Union nor the European Commission can be held responsible for them.

\bibliography{RFC}

\end{document}